\documentclass[12pt]{article}
\usepackage{authblk,cite,amssymb,amsmath}
\begin{document}
\title{Dimensional Reduction in 6D Standing Waves Braneworld}
\author{Otari Sakhelashvili \thanks{otosaxel@gmail.com}}
\affil{Javakhishvili Tbilisi State University \\
3 Chavchavadze Ave., Tbilisi 0179, Georgia}
\maketitle

\begin{abstract}

We found cosmological solution of the 6D standing wave braneworld model generated by gravity coupled to a massless scalar phantom-like field. By obtaining a full exact solution of the model we found a novel dynamical mechanism in which the anisotropic nature of the primordial metric gives rise to expansion of three spatial brane dimensions and affectively reduction of other spatial directions. This dynamical mechanism can be relevant for dimensional reduction in string and other higher dimensional theories in the attempt of getting a 4D isotropic expanding space-time.


\end{abstract}


Braneworld models with large extra dimensions \cite{Hi,brane} have been very useful in addressing several problems in high energy physics (for reviews see \cite{reviews}). Most of the braneworlds are realized as time-independent configurations, however, mostly within the framework of cosmological studies, there have been proposed models that employ time-dependent metrics and matter fields \cite{S}. Recently was proposed novel non-stationary braneworld generated by standing gravitational waves minimally coupled to a bulk scalar field \cite{Wave,6D,Oto}. In this paper we study cosmological solution to the specific 6D standing waves braneworld \cite{Oto} involved a phantom-like scalar fields. In our model the phantom scalar field is present only in the bulk and does not couple to ordinary matter on the brane (to avoid the well-known problems of stability which occur with ghost fields).

We consider 6D space-time, having the signature $(+,-,-,-,-,-)$, with a non-self interacting phantom-like scalar field coupled to gravity:
\begin{equation}\label{action}
  S = \int d^6x \sqrt g \left( \frac{M^4}{2}R - \frac 12 g^{MN}\partial _M \phi \partial _N\phi  + L_b \right)~,
\end{equation}
where $L_b$ is a brane lagrangian and $M$ is the fundamental scale, which relates to the 6D Newton constant:
\begin{equation}
G = \frac {1}{8\pi M^4}~.
\end{equation}
Capital Latin indexes numerate the coordinates of 6D space-time and we use the units where $c = \hbar = 1$.

Variation of the action (\ref{action}) with respect to $g_{AB}$ leads to the 6D Einstein equations:
\begin{equation}\label{Einstein1}
  R_{AB} = \frac {1}{M^4}\left( - \partial_A \phi \partial_B \phi + T_{AB} - \frac 14 g_{AB}T \right)~.
\end{equation}
Here $T_{AB}$ is the brane energy-stress tensor:
\begin{equation} \label{T}
  T_A^B = M^4diag[\lambda_t,\lambda_\alpha,\lambda_\alpha,\lambda_\alpha,0,\lambda_\theta]\delta(r)~, ~~~~~ T = M^4\lambda \delta(r) ~,
\end{equation}
where $\lambda = \sum \lambda_{AA}$.

To solve the system (\ref{Einstein1}) and (\ref{T}), similar to our previous article \cite{Oto}, we take the metric {\it ansatz}:
\begin{equation} \label{ansatz}
  ds^2 = e^S dt^2 - a^2e^u \left(dx^2 + dy^2 + dz^2\right) - dr^2 - \frac{1}{a^6}e^{-3u} d\theta^2~,
\end{equation}
and suppose that the metric functions, $S(|r|)$, $u(t,|r|)$, and the phantom-like scalar field, $\phi(t,|r|)$, depend only on time, $t$, and on the modulus of the extra dimension coordinate, $r$. In (\ref{ansatz}) $a = a(t)$ denotes cosmological scale factor and the second extra dimension, $\theta$, is considered to be compact, curled up to the unobservable sizes for the present accelerators energies.

By the metric (\ref{ansatz}) we want to describe geometry of the brane placed at the origin of the large extra dimension, $r$.  Advantage of the {\it ansatz} (\ref{ansatz}) is that it looks symmetric in the brane coordinates $x$, $y$ and $z$, while in 5D standing waves braneworlds \cite{Wave} the brane 3-plain is non-symmetrically warped. Also (\ref{ansatz}) is simpler than the metrics considered in the models \cite{6D}, since it contains only one singular point, $r = 0$, where the brane is placed to smooth the singularity.

The system of Einstein equations (\ref{Einstein1}) for the {\it ansatz} (\ref{ansatz}) takes the form \cite{Oto}:
\begin{eqnarray} \label{system_1}
  12 \frac{\dot{a}\dot{u}}{a} + 3\dot{u}^2 + 12 \frac{\dot{a}^2}{a^2} - \frac 12 S''e^S - \delta (r) S'e^S - \frac 14 S'^2 e^S &=& \frac {1}{M^4}\dot{\phi}^2 + e^S\left( \lambda_t - \frac 14 \lambda\right)\delta(r)~, \nonumber\\
  3u'\left(2\frac{\dot a}{a} + \dot u \right) &=& \frac {1}{M^4} \phi'\dot{\phi}~, \nonumber\\
  \frac 12 S'' + \frac 14 S'^2 + 3u'^2 + S'\delta(r) &=& \frac {1}{M^4}\phi'^2 + \frac 14 \lambda \delta(r)~, \\
  \frac {e^{-S}}{4} \left[ -4\frac{\ddot{a}}{a} - 2\ddot{u} + 4\frac{\dot{a}^2}{a^2} + S'e^Su' + 2u''e^S + 4 u'e^S\delta(r)\right] &=& - \left( \lambda_\alpha - \frac 14 \lambda \right)\delta(r)~, \nonumber\\
  \frac {3 e^{-S}}{4} \left[ - 4\frac{\ddot{a}}{a} - 2\ddot{u} + 4\frac{\dot{a}^2}{a^2} + S'e^Su' + 2u''e^S + 4 u'e^S\delta(r)\right] &=& \left( \lambda_\theta - \frac 14\lambda \right)\delta(r) ~.\nonumber
\end{eqnarray}
where overdots and primes mean derivatives with respect to $t$ and $|r|$, respectively.

In our previous paper \cite{Oto} we considered localization problem of matter fields on the brane and for simplicity assume the metric field, $u$, to be proportional to the bulk phantom-like field, $\phi$. In this paper we want to study cosmology of the model and, as in similar 5D case \cite{5D-Cos}, consider more general assumption, namely:
\begin{eqnarray} \label{u}
  u(t,r)' = \frac{1}{\sqrt {3M^4}}~\phi(t,r)'~, \nonumber\\
  \dot u(t,r) + 2\frac{\dot a}{a} = \frac{1}{\sqrt {3M^4}}~\dot\phi(t,r)~.
\end{eqnarray}
At the origin $r=0$ we introduce also the jump conditions:
\begin{eqnarray} \label{general}
  \{ S' \} = -\lambda_t + \frac14 \lambda ~, \nonumber \\
  \{ u' \} = \frac 14 (\lambda_\theta - \lambda_\alpha)~, \\
  \lambda_t = 0 ~.\nonumber
\end{eqnarray}
These relation, together with (\ref{u}) and (\ref{general}), brings the system (\ref{system_1}) to the form:
\begin{eqnarray} \label{system_2_1}
  S'' + \frac 12 S'^2 = 0 ~, \nonumber\\
  - 2\ddot{u} + S'e^Su' + 2u''e^S = 0 ~, \\
  - \frac{\ddot{a}}{a} + \frac{\dot{a}^2}{a^2} = 0~.\nonumber
\end{eqnarray}

Solution of the first equation of (\ref{system_2_1}) with the boundary condition:
\begin{equation}
  S(0) = 0~,
\end{equation}
(leading to the Minkowski metric at $r=0$) is:
\begin{equation}\label{S_sol}
  S = \ln \left(1 + \frac {|r|}{b} \right)^2~,
\end{equation}
$b$ is a constant corresponding to the width of the brane towards the large extra dimension $r$.

Using (\ref{S_sol}) and letting
\begin{equation}
  u(t,|r|) = \sin (\omega t) f(|r|)~,
\end{equation}
the second equation of (\ref{system_2_1}) takes the form:
\begin{equation}
  f'' + \frac{1}{b + |r|} f' + \frac{b^2\omega^2}{(b + |r|)^2}f = 0~.
\end{equation}
Solution to this equation, again leading to the Minkowski metric at $r=0$, is:
\begin{equation} \label{f}
  f(|r|) = C\sin \left( b\omega \ln\left[1 + \frac{|r|}{b}\right]\right) ~,
\end{equation}
where $C$ is another integration constant.

Solution of the last equation of (\ref{system_2_1}) has the form:
\begin{equation}\label{H_sol}
  a(t) = e^{H t}~,
\end{equation}
which describes the expansion law of the brane 3-surface.

To obtain the full set of solutions we need also to determine the $\lambda$ coefficients in (\ref{general}). Using (\ref{S_sol}) and (\ref{f}) the jump conditions (\ref{general}) takes the form:
\begin{eqnarray} \label{general_2}
 \{S'\} = \frac 2b ~, \nonumber\\
 \{u'\} = C \omega \sin (\omega t)~.
\end{eqnarray}
Compering (\ref{general}) and (\ref{general_2}) we find:
\begin{eqnarray} \label{general_4}
  \lambda_\theta = \frac 2b + 3C\omega \sin (\omega t)~, \nonumber\\
  \lambda_\alpha = \frac 2b - C\omega \sin (\omega t)~,
\end{eqnarray}
where the first terms are the brane tensions and the second terms correspond to the brane oscillations.

Finally the solution of Einstein equations for our {\it ansatz} (\ref{ansatz}) is:
\begin{equation} \label{s_ansatz}
  ds^2 = \left( 1 + \frac{|r|}{a}\right)^2 dt^2 - e^{2H t}e^{sin(\omega t) f(|r|)} \left(dx^2 + dy^2 + dz^2\right) - dr^2 - e^{-6H t} e^{-3 sin(\omega t) f(|r|)}d\theta^2~,
\end{equation}
where the function $f(|r|)$ is done by (\ref{f}). This expression differs from the analogous solution of \cite{Oto} by the exponential scale factors  $e^{\pm Ht}$.  So amplitudes of the oscillatory exponents in (\ref{s_ansatz}) increases/decreases with time depending on the sign of the constant $H$, wile the brane is resided at $r = 0$, in one of the nodes of the bulk standing waves, since $f(0) = 0$.

For short time intervals we can assume $Ht \to 0$ and, if the frequency of the standing waves $\omega$ is much larger than the frequencies associated with the energies of particles on the brane, one can perform a time averaging of the oscillating exponents in (\ref{s_ansatz}). In \cite{Oto,GMM} it was explicitly demonstrated that the resulting $r$-dependent functions form potential wells and can provide pure gravitational localization of matter fields on the brane.

For the large (cosmological) time intervals we can perform a time averaging of the oscillating exponents in (\ref{s_ansatz}) even for small frequencies $\omega$. Using the formula for time averages \cite{Wave,GMM}:
\begin{equation}
  \langle e^{f sin(t)}\rangle = I_0(f) ~,
\end{equation}
where $I_0(f)$ denotes the zero order modified Bessel function, the metric (\ref{s_ansatz}) reduces to:
\begin{equation} \label{metric1}
  ds^2 = \left( 1 + \frac{|r|}{a}\right)^2 dt^2 - e^{2H t}I_0(f) \left(dx^2 + dy^2 + dz^2\right) - dr^2 - e^{-6H t} I_0(-3f) d\theta^2~.
\end{equation}
The properties of the space-time (\ref{metric1}) crucially depend on the sign of the constant $H$. For the positive constant,
\begin{equation}
H > 0~,
\end{equation}
the space-time (\ref{metric1}) expands exponentially in the $x$, $y$ and $z$ directions, while the angle $\theta$ squeezes. This means that in a macroscopical time interval the space will effectively have three space-like dimensions (since $f(0) = 0$ and $I_0(0) = 1$),
\begin{equation} \label{m_ansatz}
ds^2 = dt^2 - e^{2H t} \left( dx^2 + dy^2 + dz^2 \right)~,
\end{equation}
i.e. spatial volume performs inflationary expansion.

It is known that braneworld models can provide with a geometrical mechanism of dimensional reduction supported by a curved extra dimension. For example, recently within the type IIB superstring theory it was introduced a 10-dimensional model where three out of nine spatial directions start to expand, leading to a space with $SO(3)$ symmetry instead of $SO(9)$ \cite{KEK}. In this paper we have demonstrated that a similar dynamical mechanism can be formulated within the standing wave braneworld models generated by gravity coupled to a phantom-like scalar field (see also \cite{5D-Cos}). Namely, by starting with an anisotropic 6D metric and leaving it evolve for large times, certain spatial dimensions will shrink to zero-size while others will expand in an accelerated way. Without changing of the main features of the model the number of compact extra dimensions $\theta$ in (\ref{metric1}) can be increased \cite{N-dim}. So the mechanism of dynamical dimensional reduction of multi-dimensional surfaces obtained in this paper could be useful for wide class of string models.


\section*{Acknowledgments}
I want to thank professor Merab Gogberashvili, who opened me beautiful world of braneworlds and helps me to be a physicist.

This research was supported by the grant of Shota Rustaveli National Science Foundation $\#{\rm DI}/8/6-100/12$.



\begin{thebibliography}{99}

\bibitem{Hi} N. Arkani-Hamed, S. Dimopoulos and G. Dvali,
            Phys. Lett. {\bf B 429} (1998) 263,
            arXiv: hep-ph/9803315; \\
             I. Antoniadis, N. Arkani-Hamed, S. Dimopoulos and G. Dvali,
            Phys. Lett. {\bf B 436} (1998) 257,
            arXiv: hep-ph/9804398.

\bibitem{brane} M. Gogberashvili,
               Int. J. Mod. Phys. {\bf D 11} (2002) 1635,
               arXiv: hep-ph/9812296;
               Mod. Phys. Lett. {\bf A 14} (1999) 2025,
               arXiv: hep-ph/9904383; \\
                L. Randall and R. Sundrum,
               Phys. Rev. Lett. {\bf 83} (1999) 3370,
               arXiv: hep-ph/9905221;
               Phys. Rev. Lett. {\bf 83} (1999) 4690,
               arXiv: hep-th/9906064.


\bibitem{reviews} V.A. Rubakov,
                Phys. Usp. {\bf 44} (2001) 871 (Usp. Fiz. Nauk {\bf 171} (2001) 913);\\
                 D. Langlois,
                Prog. Theor. Phys. Suppl. {\bf 148} (2003) 181,
                arXiv: hep-th/0209261;\\
                 P.D. Mannheim,
                {\it Brane-localized Gravity} (World Scientific, Singapore 2005); \\
                 R. Maartens and K. Koyama,
                Living Rev. Rel. {\bf 13} (2010) 5,
                arXiv: 1004.3962 [hep-th].

\bibitem{S}  M. Gutperle and A. Strominger,
            JHEP {\bf 0204} (2002) 018,
            arXiv: hep-th/0202210; \\
             M. Kruczenski, R.C. Myers and A.W. Peet,
            JHEP {\bf 0205} (2002) 039,
            arXiv: hep-th/0204144; \\
             V.D. Ivashchuk and D. Singleton,
            JHEP {\bf 0410} (2004) 061,
            arXiv: hep-th/0407224; \\
             C.P. Burgess, F. Quevedo, R. Rabadan, G. Tasinato and I. Zavala,
            JCAP {\bf 0402} (2004) 008,
            arXiv: hep-th/0310122.

\bibitem{Wave} M. Gogberashvili and D. Singleton,
              Mod. Phys. Lett. {\bf A 25} (2010) 2131,
              arXiv: 0904.2828 [hep-th]; \\
               M. Gogberashvili, P. Midodashvili and L. Midodashvili,
              Int. J. Mod. Phys. {\bf D 21} (2012) 1250081,
              arXiv: 1209.3815 [hep-th]; \\
               M. Gogberashvili,
              JHEP {\bf 09} (2012) 056,
              arXiv: 1204.2448 [hep-th].

\bibitem{6D} L.J.S. Sousa, W.T. Cruz and C.A.S. Almeida,
            Phys. Rev. {\bf D 89} (2014) 064006, arXiv: 1311.5848 [hep-th];\\
             P. Midodashvili,
            Int. J. Theor. Phys. {\bf 53} (2014) 1174, arXiv: 1211.0206 [hep-th].

\bibitem{Oto} O. Sakhelashvili,
             Int. J. Theor. Phys. {\bf 53} (2014) 1940, arXiv: 1311.1030 [gr-qc].

\bibitem{GMM} M. Gogberashvili, P. Midodashvili and L. Midodashvili,
            Phys. Lett. {\bf B 702} (2011) 276, arXiv: 1105.1701 [hep-th];
            Phys. Lett. {\bf B 707} (2012) 169, arXiv:1105.1866 [hep-th]; \\
             M. Gogberashvili, O. Sakhelashvili and G. Tukhashvili,
            Mod. Phys. Lett. {\bf A 28} (2013) 1350092, arXiv: 1304.6079 [hep-th].

\bibitem{KEK} S.-W. Kim, J. Nishimura and A. Tsuchiya,
             Phys. Rev. Lett. {\bf 108} (2012) 011601, arXiv: 1108.1540 [hep-th];
             Phys. Rev. {\bf D 86} (2012) 027901, arXiv: 1110.4803 [hep-th].

\bibitem{5D-Cos}  M. Gogberashvili, A. Herrera-Aguilar and D. Malagon-Morejon,
              Class. Quantum Grav. {\bf 29} (2012) 025007, arXiv: 1012.4534 [hep-th]; \\
               M. Gogberashvili, A. Herrera-Aguilar, D. Malagon-Morejon and R.R. Mora-Luna,
              Phys. Lett. {\bf B 725} (2013) 208, arXiv: 1202.1608 [hep-th].

\bibitem{N-dim} M. Gogberashvili, P. Midodashvili, G. Tukhashvili,
              Gen. Rel. Grav. {\bf 46} (2014) 1697, arXiv: 1310.5696 [gr-qc].

\end{thebibliography}
\end{document}